\documentclass{article}

\usepackage{arxiv}

\usepackage[utf8]{inputenc} 
\usepackage[T1]{fontenc}    
\usepackage{hyperref}       
\usepackage{url}            
\usepackage{booktabs}     
\usepackage{amsfonts}       
\usepackage{nicefrac}       
\usepackage{microtype}      
\usepackage{lipsum}
\usepackage{algorithm}
\usepackage{algpseudocode}
\usepackage{mathtools}

\newcommand{\Neighmath}{\mathcal{N}}

\newcommand{\Uniqgen}{U}
\newcommand{\Uniqneighmathlat}{\Uniqgen_{\Neighmath}}
\newcommand{\Uniqneighmath}{\Uniqgen_{\Neighmath}}
\newcommand{\Uniqneigh}{$\Uniqneighmathlat\:$}

\newcommand{\degmono}{k}

\newcommand{\Uniqdegreesinglemath}{\Uniqgen_{\degmono}^{}}

\title{Privacy and Uniqueness of Neighborhoods\\ in Social Networks}

\author{
  Daniele Romanini \\
  Department of Computer Science, Aalto University, Espoo, Finland\\
  DTU Compute, Technical University of Denmark, Lyngby, Denmark\\
  \texttt{daler.romanini@gmail.com} \\
   \And
 Sune Lehmann \\
DTU Compute, Technical University of Denmark, Lyngby, Denmark\\
  \texttt{sljo@dtu.dk} \\
     \And
    Mikko Kivel{\"a}\\
  Department of Computer Science, Aalto University, Espoo, Finland\\
  \texttt{mikko.kivela@aalto.fi} \\
}

\begin{document}
\maketitle

\begin{abstract}
The ability to share social network data at the level of individual connections is beneficial to science: not only for reproducing results, but also for researchers who may wish to use it for purposes not foreseen by the data releaser. 
Sharing such data, however, can lead to serious privacy issues, because individuals could be re-identified, not only based on possible nodes’ attributes, but also from the structure of the network around them.
The risk associated with re-identification can be measured and it is more serious in some networks than in others. 
Various optimization algorithms have been proposed to anonymize the network while keeping the number of changes minimal. 
However, existing algorithms do not provide guarantees on where the changes will be made, making it difficult to quantify their effect on various measures. 
Using network models and real data, we show that the average degree of networks is a crucial parameter for the severity of re-identification risk from nodes' neighborhoods. 
Dense networks are more at risk, and, apart from a small band of average degree values, either almost all nodes are re-identifiable or they are all safe. 
Our results allow researchers to assess the privacy risk based on a small number of network statistics which are available even before the data is collected. 
As a rule-of-thumb, the privacy risks are high if the average degree is above 10. 
Guided by these results we propose a simple method based on edge sampling to mitigate the re-identification risk of nodes. 
Our method can be implemented already at the data collection phase. 
Its effect on various network measures can be estimated and corrected using sampling theory. 
These properties are in contrast with previous methods arbitrarily biasing the data. 
In this sense, our work could help in sharing network data in a statistically tractable way.
\end{abstract}


\section{Introduction}\label{intro}
Much of the complexity of social systems, ranging from individual social groups to whole societies, is encoded in the structure of the interactions between individuals. 
These \emph{social networks} are a focal point in quantitative research aiming to explain how social systems work from the microscopic level of individuals to the macroscopic level with millions of people.  
This area of research spans studies from the beginning of the previous century \cite{scott1988social} to more recent explosion of work on automatically collected digital traces of up to millions of individuals. The topics of this research are numerous and are related to many of the burning issues in the world such as opinion polarisation \cite{garimella2018quantifying, baumann2020modeling, chen2020polarization}, disease spreading \cite{barrat2008dynamical, firth2020combining, barrat2020effect}, and social segregation \cite{leo2016socioeconomic, lee2019homophily, asikainen2020cumulative}, to name a few. 
What is common to this wide range of research is that it is all based on data describing the structure of social networks, represented simply through nodes (individuals) and edges (their connections). 

In order for the scientific community working on social networks and related problems to function efficiently, it is imperative that researchers can share social network data. 
First, only sharing network data allows scientists to reproduce already presented results and validate (or rebut) them. 
This process is the basis of the scientific method, and essential in order to avoid a crisis in reproducing scientific results which have already been reported in multiple fields \cite{baker20161, Hutson2018Artificial, miyakawa2020no}. 
Second, social network data can be (and is routinely) reused to answer questions that were not conceived by the original research group releasing the data. 
In some cases the social network data itself could turn out to be more important for the scientific progress than the answer related to the original research question that was answered in the article releasing the data.

Social network data can contain sensitive information which can make it impossible to share such data due to privacy concerns. 
For example, the structure of the network itself, i.e.~who connects to whom, can be sensitive -- or the individuals could have sensitive attributes attached to them. 
As networks contain complex multidimensional information compared to, for instance, tabular data, they result in a complex problem of how to share data in a privacy preserving way. 

Many of the data anonymization techniques developed over the years are simply unsuitable for social network data.
A specialised set of network-based techniques have been developed to address this problem. 
Within these there are a number of different scenarios (threat models) of how exactly a malicious party might be able to deanonymize the network, and even greater number of methods to anonymize network data (see Section~\ref{privacy}). 
The focus in the literature has been on developing methods that minimize the changes to the networks -- or particular network statistics -- while making them less vulnerable to a particular deanonymization attack. 

In this paper we analyse the factors that make networks vulnerable in the first place. 
We do this by studying different network models, varying their parameters 
and measuring their vulnerability in various configurations.
A better understanding of the causes of vulnerability can both guide to a better design of solutions to address each vulnerability and, more importantly, lead to a more accurate risk assessment. 
Based on our findings, we explore a simple anonymization strategy based on random sampling. 
Here, the idea is that changes to the networks can be substantial as long as they are statistically tractable such that they can be corrected for further analysis. 

We focus on a single well-studied threat model known as the neighborhood attack \cite{zhou2008preserving}. 
In this scenario the attacker has knowledge of the structure of the network neighborhood of a victim node, which is represented by the number of a person's friends and the connections between them. 
Based on this information, the attacker tries to re-identify the victim in the anonymized network. 
If a node is re-identified in this way, the attacker could get access to private information potentially attached to the node, or the wider position of the node in the network. 
Furthermore, the attacker could also re-identify the target's connections (e.g. friends or relatives). 
The latter case could be possible also if an individual who knows the structure of their own neighborhood is able to recognize himself in a certain dataset. 
In this case, then they would potentially be able to re-identify their friends and their connections. 
The links themselves, even without explicit labels, could be already a private information (as they represent, for example, the people we communicate or spend time with), and thus important to protect.

The remainder of this paper is organized as follows. 
In Section \ref{privacy} we give an overview of the already existing works on privacy in social networks, in particular on neighborhood attack. 
In Section \ref{approach} we formally introduce the threat model we use and the related network diagnostics. 
We present our main findings in Sections \ref{uniquenessrandom} and \ref{uniquenessreal}, showing how network models can also be a good proxy for real-world networks with respect to uniqueness of neighborhoods. 
Finally, in Section \ref{anonymizingnet}, we show how our findings can be applied in practice to lower the re-identification risk in networks.

\section{Privacy in Social Networks}\label{privacy}

Privacy in data sharing is a growing area of research. 
With increasing amount of personal data and data analysis techniques, the risk for users' privacy has dramatically increased. 
Existing privacy-preserving and anonymization methods often rely on different definitions of privacy that have been mainly developed for tabular data, even though there has been some efforts to extend those definitions to networks.

Some of the most popular existing definitions of privacy are: \textit{na\"{i}ve anonymization} or \textit{pseudonymization}, which consists in dropping the entities' labels, and replacing them with random labels; 
\textit{random perturbation/noise injection} \cite{kargupta2003privacy}; 
\textit{k-anonymity} \cite{sweeney2002k}, where the dataset is modified such that each entry is indistinguishable from at least other $k-1$ entries (or, in other words, each equivalence class contains at least $k$ values), reducing the re-identification attack success probability to $\frac{1}{k}$; 
\textit{$\ell$-diversity} and \textit{t-closeness}, which are group-based anonymization techniques for labelled data, developed to strengthen the definition of privacy given by \textit{k-anonymity}; 
\textit{Differential privacy} \cite{Dwork:2006:DP:2097282.2097284, dwork2008differential, dwork2014algorithmic}, a mathematical definition used to develop algorithms to query data, ensuring the privacy of the response, or, more recently, to learn generative models for private data sharing \cite{zhang2017privbayes}. 
Some differential privacy methods have been developed to perform specific network analysis tasks \cite{hay2009accurate, lu2014exponential, shen2013mining}, aiming to protect only some information, such as the ones related to nodes (node differential privacy \cite{kasiviswanathan2013analyzing}) or edges (edge differential privacy). 

In networks, in addition to node-attributes, there is a further threat due to the network structure itself. 
That is, a node or a link can be identifiable by its ``location'' in the network.
In fact, without this consideration, one could think that a na\"{i}ve anonymization approach could be enough for sharing network data. 
To account for these structural threats, the private network components, such as nodes or links, need to be structurally indistinguishable in the network as a whole. 
To address these type of general structural attacks, concepts such as \textit{k-automorphism} \cite{zou2009k} and \textit{k-isomorphism} \cite{cheng2010k} have been developed. These assess the privacy risks arising from an the attacker with knowledge of the whole graph or any part of it.

An alternative to anonymizing the whole network structure is to compute an array of network statistics, anonymize the array, and either share the statistics or uniformly sample a network with the set of anonymized statistics. 
One such approach is to share the block matrix of a stochastic block model \cite{hay2008resisting}.
Further, some papers have proposed sharing the 
$dK$-series 
\cite{sala2011sharing} or a hierarchical random graph \cite{xiao2014differentially} under differential privacy. Differential privacy is certainly the state-of-the-art in data privacy research, and those studies proposed promising directions for private data-sharing.
However, these statistics-based methods have the serious drawback that they only retain the particular statistics that are measured; or in the case of models, structure that is encoded in each model. 
For example, the $dK$-series is often used to retain degree correlations ($dK-2$-series) (and possibly triangle counts) \cite{horawalavithana2019privacy}, but disregard any mesoscopic structures, such as a particular community structure. 
Further, depending on the statistics, it can be difficult to develop algorithms which are guaranteed to sample uniformly from networks with specific set of statistics \cite{orsini2015quantifying}. 

An attacker does not need to have a knowledge of the full network, but some local structural features can make a node unique and thus re-identifiable in a network. 
Typical local features like this include the number of connections (degree) or the structure of its neighborhood. 
The neighborhood attack, which consists in the re-identification of a node based on its neighborhood, was first introduced in \cite{zhou2008preserving}. 
This work focused on the 1-hop neighborhood of a node, which is composed by its immediate neighbors only, in unweighted, undirected networks.
Related concepts such as 
\textit{k-degree anonymity} \cite{liu2008towards} and \textit{k-neighborhood anonymity} \cite{zhou2008preserving}, were developed to asses the risk of attacks such as the \textit{neighborhood attack}. 

Several algorithms have been developed for making networks safe from neighborhood attacks, while changing the network a minimal amount.
Finding an optimal solution (e.g. adding the minimum amount of edges) to reach the anonymity is a NP-Hard problem \cite{chester2013complexity}. 
The heuristic that \cite{zhou2008preserving} adopts to anonymize two neighborhoods consists in matching neighborhood's components pairs (starting with the ones with the highest amount of vertices) and making them isomorphic if they are not already, by adding edges. 
The choice of neighborhoods pair from the two networks that need to be anonymized is done by matching the degree of the central node. 
If no matching pairs are found, then the pair with the minimum anonymization cost (computed by taking into account the amount of edges that needs to be added) is chosen. 
Ref.~\cite{tripathy2010new} proposed an alternative to this approach in terms of matching the neighborhood components, representing them with an adjacency matrix, making it extensible to more than 1-hop neighborhood. 
Focusing on the number of edges that are changed during the anonymization is not the only possible approach, but, depending on the application, one can also focus on specific network metrics. 
For example, distances between pairs of nodes have been used as such metric~\cite{okada2014k}.

Several network datasets come with some additional information on the nodes or the edges in addition to the network structure.
In fact, studies have shown that attributes and meta-data can be crucial to identify individuals in some datasets, such as locations visited, metadata associated to social media usage, and credit card transactions \cite{de2015unique, de2013unique, perez2018you}.
Neighborhood anonymization methods have been developed for anonymizing labelled neighborhoods \cite{zhou2011k,wang2014resisting}, in an attack where the attacker has also a background knowledge of the labels of the nodes part of the target's neighborhood, besides of its structure. 
Ref.~\cite{liu2015k} treats the problem of neighborhood anonymization in weighted unlabelled networks, anonymizing the network by edge addition and weight modification.

Network anonymization algorithms which modify networks by addition or removal of nodes and links aim to minimize such changes. 
In general, there is always a balance between utility and anonymity of the resulting network. 
Depending on the data, with high privacy requirements the amount of changes to the network can be very significant. 
Further, while the anonymizer knows where exactly these changes are and can evaluate their impact to specific network metrics, the end user of the anonymized data must assume that the changes could be in arbitrary places in the network. 
This makes it difficult for users to evaluate the errors or make the error estimates very high even for moderate amount of changes. 
For example, the impact to performance on classification methods such as Graph Neural Networks has been show to be significant when the worst-case places for node or edge additions are assumed \cite{sun2018adversarial, wang2018attack, zugner2018adversarial}. 

The literature on network privacy and anonymity has focused on developing algorithms that ensure various notions of anonymity while minimizing particular types of changes to networks. 
Much less work has focused on understanding the factors that cause networks to be anonymous in the first place.
For example, given a threat model, such as neighborhood attack, it would be useful to know how difficult the anonymization problem is, and
``how far" the original network is from having zero unique neighborhoods. 
In fact, if the number of unique neighborhoods is very high, it might be not worth-while to anonymize and share the network at all, as too many modifications would be required, significantly lowering the data utility. 
Ideally we would also like to have an (approximate) understanding of this difficulty without seeing the full network data.

From the existing literature we understand theoretically how the the knowledge of nodes degree, degrees of neighbors, degrees of neighbors neighbors and so on, can be used to deanonymize nodes in Erd\H{o}s-R\'enyi networks at the limit of infinite size \cite{hay2008resisting}. 
In these large networks, the degree is not enough to uniquely identify a node, but, for successive higher-order degrees, uniqueness depends on the network density. 
Further, individual instances of power-law networks and lattices have been analysed with the same approach \cite{hay2008resisting}.
Our current work is similar in nature to this previous study, but we assume the more common neighborhood attack scenario where the attacker has knowledge about the entire target's neighborhood structure. 
Further we take into account finite size Erd\H{o}s-R\'enyi (ER) networks, analyzing simultaneously both size and density, but focusing on sparse networks, as real social networks are typically sparse. 
As sparse ER networks are different from real social networks, since almost completely void of local structure, we also analyze the Watts-Strogatz and the Random Geometric Graph models which are minimal models containing such structure.

\section{Measuring privacy risks with uniqueness of neighborhoods}\label{approach}

We want to study the uniqueness of neighborhoods in networks, to understand the factors that influence uniqueness in different settings and, at the same time, shed light on which network properties are relevant to quantify the privacy risk. 
The neighborhood of a node $v$ consists of both the nodes which are adjacent to it and the links between them.
Formally, we define the neighborhood, $\mathcal{N} (v)$, to be the induced subgraph of its neighbors. 
That is, given a graph $G=(V,E)$, $\mathcal{N} (v) = (V(v),E(v))$, where $V(v)=\{(u \in V | (v,u) \in E\}$ and $E(v) = \{(u,w) \in E | u,w \in V(v)\}$.
A neighborhood of a node is unique if there are no other node neighborhoods in the network with the same graph structure (disregarding the node labels). 
Formally this means that $\mathcal{N} (v)$ is unique if there is no other node $u \in V$ for which the neighborhoods are graph isomorphic, $\mathcal{N} (v) \cong \mathcal{N} (u)$.

The uniqueness of a neighborhood guarantees that an attacker equipped with the knowledge of the neighborhood of a node can identify with absolute certainty the given node. 
However, a stronger notion of privacy can be achieved if there are multiple neighborhoods with exactly the same structure.

 We define the 
 \textit{occurrence frequency} 
 $O_{\mathcal{N} (v)}$ as the number of neighborhoods in $G$ that are isomorphic to $\mathcal{N}(v)$: 
\begin{equation} 
    O_{\mathcal{N} (v)} = \sum_{u \in V}{\delta(\mathcal{N} (v) \cong \mathcal{N} (u) )}    \,,
    \label{eq:occfreq}
\end{equation}
where
\begin{equation}\label{delta}
    \delta(\mathcal{N} (v) \cong \mathcal{N} (v')) = \begin{cases}   1, & \:if\:\mathcal{N} (v) \cong \mathcal{N} (v')\\
                            0, &  \:otherwise
            \end{cases} \,.
\end{equation}

In order to quantify the overall privacy of a network, we define the \textit{uniqueness of neighborhoods} $U_{\mathcal{N}}$ (or, simply, \textit{uniqueness}, in our case) as the fraction of unique nodes in a network:
\begin{equation}\label{unequations}
    U_{\mathcal{N}} = \sum_{v \in V}{\frac{\delta(O_{\mathcal{N} (v)} = 1)}{|V|}}    \,,
\end{equation}
where:
\begin{equation}\label{deltaunequations}
    \delta(O_{\mathcal{N} (v)} = 1) = \begin{cases}   1, & \:if\:O_{\mathcal{N} (v)} = 1\\
                            0, &  \:otherwise
            \end{cases} \,.
\end{equation}

If the value of uniqueness is equal to one (maximum uniqueness), it means that there are only unique neighborhoods in the network, thus no neighborhood is isomorphic to any other. 
Conversely, if $\Uniqneighmath=0$ (minimum uniqueness), every neighborhood occurs at least two times in $G$, and if $\Uniqneighmath=0.5$, half of the neighborhoods occur just one time in $G$.

A node could be uniquely identifiable not only by its neighborhood, but also by its degree. 
We define the \textit{degree uniqueness} $U_{k}$ of a network $G$, as the fraction of nodes in $G$ that have unique degree. 
Equations \ref{eq:occfreq}, \ref{delta}, \ref{unequations} and \ref{deltaunequations} can still be applied to the this type of uniqueness, by substituting the neighborhood $\mathcal{N}(v)$ with a function $k(v)$ which returns the degree of node $v$. 

Clearly, $U_{\mathcal{N}} \geq U_{k}$, because if a node has a unique degree, it also has a unique neighborhood. 
The two notions of uniqueness can differ from each other when there is at least one edge between the neighbors of the central node. 
Each edge in a neighborhood corresponds to a triangle where the central node is participating, and, 
in Section \ref{uniquenessrandom}, we discuss how the presence of triangles affects the value of \Uniqneigh{} and its difference with $U_{k}$. 
For this reason, we introduce the notation for the fraction of nodes with degree $k$ having at least one triangle in their neighborhood as $p_{k}$, and the expected
fraction of neighborhoods with at least one triangle as $\mathcal{N}_{\bigtriangleup}$.

In the following, we study structural privacy, and, in particular, neighborhood anonymity, in unlabelled, unweighted, undirected networks with no self-loops. 
We study the case of 1-hop neighborhoods only, but our methodology can be adapted to higher order neighborhoods or to other structures that can uniquely characterize a node, for instance the degree, or labelled neighborhoods.

\section{Uniqueness of neighborhoods in random network models}\label{uniquenessrandom}
In order to understand how networks behave in terms of uniqueness of neighborhoods, and thus vulnerability to neighborhood attacks and  difficulty of the anonymization problem, 
we study three different network models. 
The models we study are Erd\H{o}s-R\'enyi (ER) \cite{erdos1960evolution}, the Watts-Strogatz (WS) \cite{watts1998collective} and a Random Geometric Graph  (RGG) \cite{penrose2003random} model. 
We choose those models because we want to represent different levels and aspects of randomness and local structure. 
All models are specified using two parameters; here we choose to represent them with the size and average degree. 
We note that there are of course several other models for social networks (and other networks), which we have left out, and which can produce a rich set of structures such as fat-tailed degree distributions, homophily, community structure, core-periphery structure \cite{lee2019homophily, asikainen2020cumulative,toivonen2009comparative, snijders2011statistical}.

The ER model generates networks with nodes that are randomly connected by an edge with probability $p$. 
In large and sparse ER networks almost all neighborhoods are empty, and, as we explain in Section \ref{deguniqer}, there is a parameter range where ER networks are unlikely to be useful proxies for estimating uniqueness for real-world networks, since the latter typically have important local structure. 
Instead, the WS and the RGG can be used to generate networks with realistic neighborhood densities (i.e., clustering coefficient values). 
The WS model has a parameter $\beta$ representing the probability of rewiring each edge from a regular lattice structure (if $\beta = 1$, the generated graph is a random graph similar to a ER graph). 
The RGG (in its soft version, which we use) is constructed by randomly placing $n$ nodes in an Euclidean space uniformly. 
If two nodes are within a given radius $r$, they are connected with a specified probability (in our case, an exponential distribution). 
The expected average degree of RGG is roughly $\langle k \rangle \approx \pi (n-1) r^2  $. 
The WS and RGG models are not as easily amenable to derivations of explicit equations for uniqueness as the ER model, and instead, in Section \ref{uniquenessmaps}, we show the uniqueness values for suitable parts of the parameter space of all of the models.

\subsection{Degree and neighborhood uniqueness in Erd\H{o}s-R\'enyi networks}\label{deguniqer}

\begin{figure}[h!]
    \centering
    \includegraphics[]{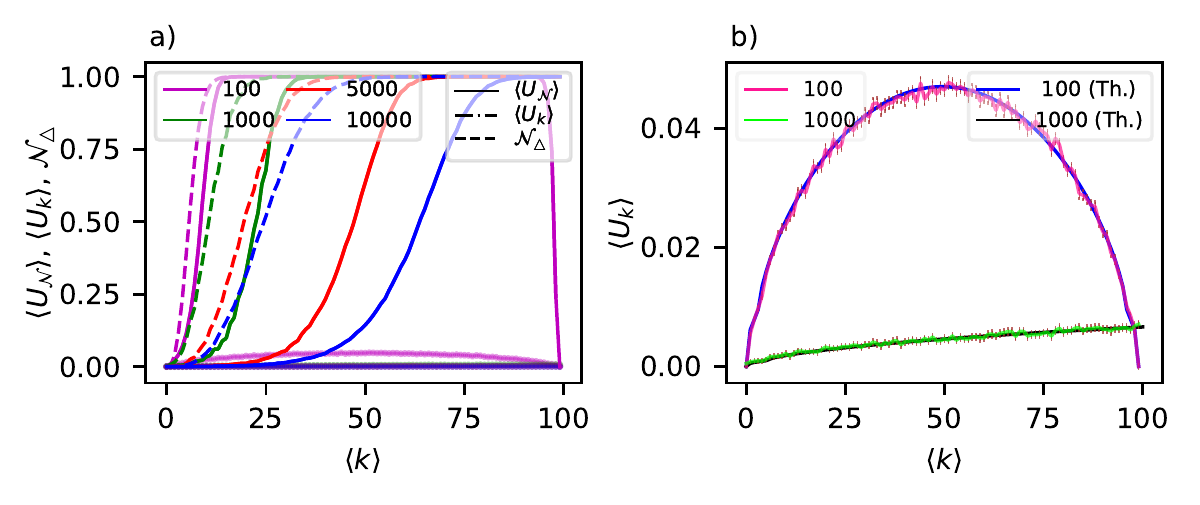}
    \caption {Neighborhood uniqueness, degree uniqueness and non-empty neighborhoods in ER networks with varying size and average degree.
    (\textit{a}) The expected uniqueness of neighborhoods $\langle U_{\mathcal{N}} \rangle$, degree uniqueness $\langle U_{k} \rangle$ and number of neighborhoods $\mathcal{N}_\Delta$ with at least one triangle (or, non-empty neighborhoods) in ER networks of size 100, 1000, 5000 and 10000 (computed as the mean of 10 independent network realizations), for values of average degree from 0 to 100. The degree uniqueness values are very close to $0$ and for network sizes from 1000 to 10000 they are below $0.02$ and covered by each other. (\textit{b}) The values of degree uniqueness for networks of size 100 and 1000 both from simulations, and the theoretical line, computed with Eq.~\ref{singlelayeruniquedegrees}. The means and errors (computed as standard error of the mean) are computed over 400 independently simulated networks for the networks with 100 nodes, and 20 independently simulated networks for the ones with 1000 nodes. This is because those are bigger systems, thus self-averaging.}
    \label{fig:ER_uniqueness}
\end{figure}

We start our investigation by analyzing the Erd\H{o}s-R\'enyi (ER) model. 
The ER is arguably the model, with adjustable size and average degree, that has the minimum number of assumptions \cite{park2004statistical}, and it has been analysed previously at the limit of infinite network size under a different attack model \cite{hay2008resisting}.
As shown in the examples of uniqueness as a function of degree for networks of sizes 100, 1000, 5000 and 10000 in Fig.~\ref{fig:ER_uniqueness}a, uniqueness in the ER networks is sensitive to both size and average degree. 
Simply analysing the limiting behavior in the size is not sufficient. 

When computing the neighborhood uniqueness in ER networks as a function of average degree, the uniqueness is at first a monotonically increasing function until it hits the maximum value ($U_{\mathcal{N}}=1$). 
On the other hand, however, when the network is complete, all the nodes are connected to each other, thus all the neighborhoods have equal structure, and there is no uniqueness. 
The neighborhoods' uniqueness remains stable for almost all the possible values of average degree, and it starts decreasing only when the network is almost complete (see Fig.~\ref{fig:ER_uniqueness}a for an illustration for networks with 100 nodes, and Fig.~\ref{fig:uniqcurvews_appendix}a for a comparison with networks with 200 and 300 nodes). 
Overall, this means that, in order to understand the behaviour of the uniqueness of neighborhoods, we can focus on small average degree values as long as we observe the transition from zero uniqueness to full uniqueness. 

As seen in Fig.~\ref{fig:ER_uniqueness}a, the larger an ER network, the larger values of average degrees are needed to observe the transitions from zero uniqueness to full uniqueness. 
It turns out that this behavior can be understood in the case of ER networks, due to the fact that the larger the ER network is, the smaller the network density $p$ for a given average degree would be. 
The overall network density is exactly the same as the expected neighborhood density, which means that most of the neighborhoods remain empty for low values of $p$ \cite{watts1998collective}.

If, in the neighborhood of a node, there are no edges between the neighbors, that neighborhood is entirely described by the degree of the node. 
Further, if all the nodes have empty neighborhoods ($\mathcal{N}_\Delta=0$) the neighborhood uniqueness  ($U_{\mathcal{N}}$) is equal to the degree uniqueness ($U_{k}$).
In fact, this is exactly what we seen in Fig.~\ref{fig:ER_uniqueness}a: transition from empty to non-empty neighborhoods happens before the neighborhood uniqueness transitions. 
In addition, the degree uniqueness values remain very low, especially for larger networks. 
This means that, before the neighborhood uniqueness transition, the neighborhood sizes (i.e., degrees) alone are not enough to make the nodes unique. The empty neighborhoods need to contain at least small amount of edges in order for the neighborhood uniqueness transition to take place.
When this behavior is combined with the fact that smaller networks have denser neighborhoods, we can understand why the uniqueness transition is driven by the network size.

In ER networks we do not only need to resort to simulating networks for particular parameter values, but we can also derive the formulas for both $\langle U_{k} \rangle$ and $\mathcal{N}_\Delta$.
For $\langle U_{k} \rangle$ we can write down the probability that a node has degree $k$ and no other node has the same degree, and sum over all possible degrees:
\begin{equation}\label{singlelayeruniquedegrees}
    \langle \Uniqdegreesinglemath \rangle = \sum_{k=0}^{n-1} p_k (1 - p_k)^{n-1} \,,
\end{equation}
where $p_k$ is the probability of a node to have degree $k$, according to the degree distribution of ER networks (binomial distribution):
\begin{equation}
\begin{split}
    p_k\: &= \binom{n-1}{k}p^{k} (1-p)^{n-1-k} \\
        &= \frac{(n-1)!}{k! (n-1-k!)} \bigg(\frac{\langle k \rangle}{n-1}\bigg)^{k} \bigg(1- \frac{\langle k \rangle}{n-1}\bigg)^{n-1-k} \,.
\end{split}
\end{equation}

Eq.~\ref{singlelayeruniquedegrees} matches the simulations as seen in Fig.~\ref{fig:ER_uniqueness}b. 
We can also see that the \textit{degree uniqueness} curve has a convex shape, and the maximum fraction of nodes with unique degrees is reached when the average degree is half of the maximum one ($n - 1$). 
We can confirm this in general by taking the first derivative of ${\langle \Uniqdegreesinglemath \rangle}$, and evaluating it at $\langle k \rangle = \frac{n - 1}{2}$, with the result being equal to 0:
\begin{equation}
     \frac{d \: {\langle \Uniqdegreesinglemath \rangle}
    }{d k}\bigg|_{{\langle k \rangle}=\frac{n-1}{2}} = 0 \,.
\end{equation}

To compute $\mathcal{N}_\Delta$, we first compute the probability that the neighborhood of a node with degree $k$ is non-empty:

\begin{equation}\label{singletriangleer}
    \mathcal{N}_\Delta(k)
    = 1 - (1-p)^{\binom{k}{2}} = 1 - (1-p)^{\frac{k(k-1)}{2}} \,.
\end{equation}
The expected fraction of non-empty neighborhoods is then given by taking the expectation over the degree distribution,
\begin{equation}\label{totaltriangles}
    \mathcal{N}_\Delta
    = \sum_{k=0}^{n-1} \mathcal{N}_\Delta(k) p_k \,.
\end{equation}

The uniqueness of neighborhoods in finite ER networks is driven by the sparsity of the neighborhoods, which is sensitive to the network size for a given average degree. 
As real-world (social) networks often show significantly different neighborhood density from random networks \cite{watts1998collective}, using ER networks as a model for them will likely underestimate the privacy risks. 
This reasoning is likely valid also for other random networks which are known to show a tree-like structure (such as sparse configuration models).

\subsection{Uniqueness maps}\label{uniquenessmaps}

\begin{figure}[h!]
\centering
  \advance\leftskip-0.2cm
  \includegraphics[]{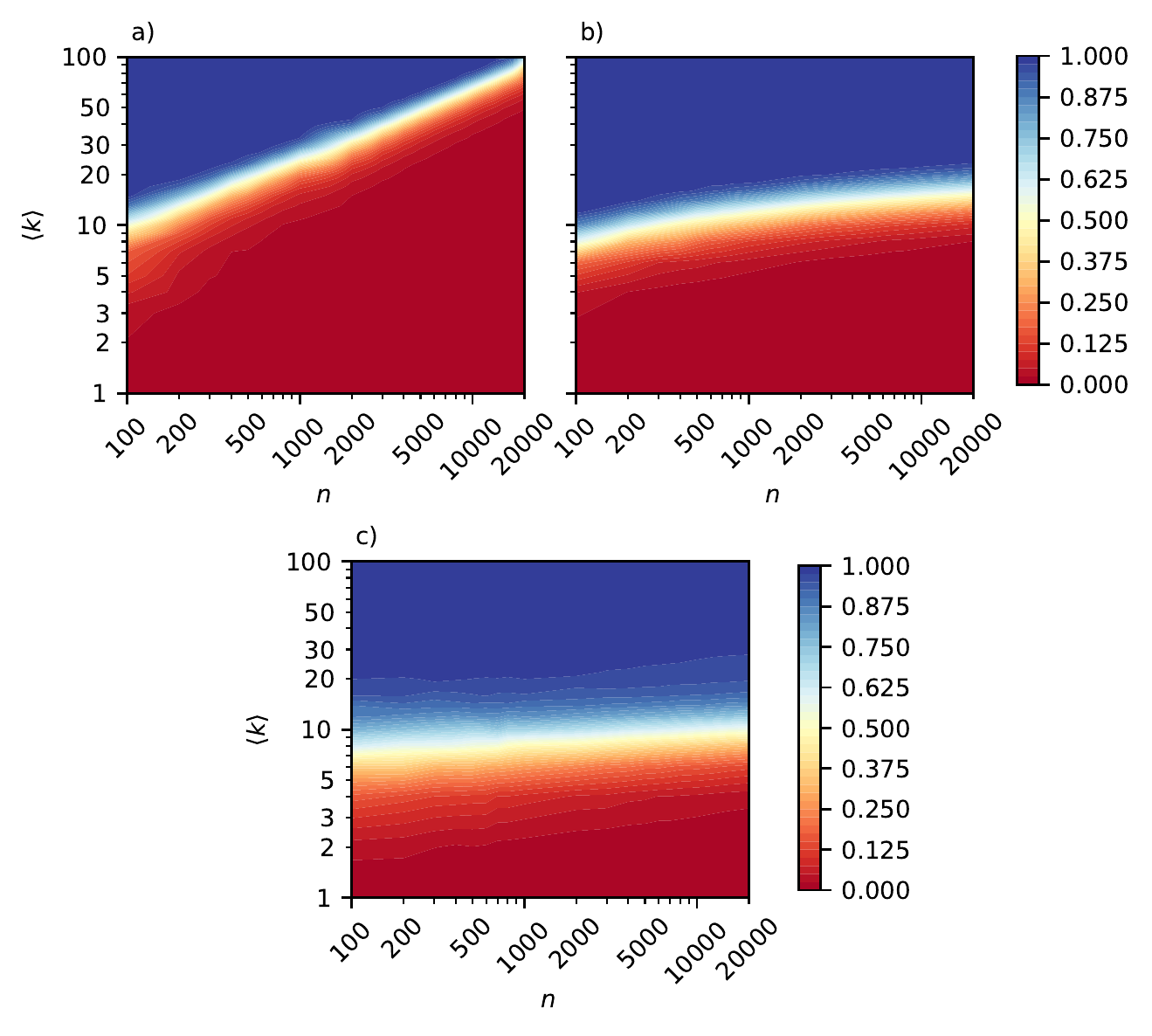}
\caption{Uniqueness maps: heatmaps representing the variation of the uniqueness of neighborhoods value (in color: the blue area corresponds to a uniqueness value equal to 1, while the red area corresponds to a uniqueness value equal to 0) in networks generated according to ER (\textit{a}), WS with probability of rewiring $\beta = 0.5$ (\textit{b}) and RGG (\textit{c}) models, in function of the average degree $\langle k \rangle$ (on the vertical axis) and network size $n$ (on the horizontal axis). The axis are in logarithmic scale. The uniqueness values are computed as the mean of 10 independently simulated networks for each average degree and network size value. Note that, even if not clearly shown in this figure, in networks with 100 nodes, the maximum possible value of average degree is 99, where $\Uniqneighmath = 0$ (see Fig.~\ref{fig:ER_uniqueness} for an example of the uniqueness behaviour in ER networks with 100 nodes).}\label{fig:uniquenessmaps}
\end{figure}

We construct \textit{uniqueness maps}, which show the value of the uniqueness of neighborhoods as a function of both the average degree and the network size. 
In addition to ER networks, we include in our analysis models with neighborhood structure which is dense and independent of the network size (a property which is more typical of real-world social networks). 
We build \textit{uniqueness maps} for ER, WS (with $\beta = 0.5$) and RGG models, with number of nodes from 100 to 20000 and average degree from 1 to 100. 
The maps of the three models can be seen in Fig.~\ref{fig:uniquenessmaps}.

We can see that the area where both anonymous nodes and identifiable nodes exists simultaneously is a small band running across each heatmap. 
That is, nodes in those networks are mostly either almost fully anonymous, or vulnerable to re-identification, and the transition between these two states happens over a relatively small range of average degree values. 
The area where networks are almost fully anonymous is the ones with low values of average degree, and we can see that the number of anonymous nodes increases more slowly as more local structure is present in the network. 
In fact, for random networks such as ER (Fig.~\ref{fig:uniquenessmaps}a), the uniqueness of neighborhoods is significantly lower compared to the other models when the number of nodes is high -- with even higher values of average degree. 
In ER networks, at the limit of infinitely large networks and a given average degree, any neighborhood is almost surely empty and we can find two nodes that have structurally equivalent neighborhoods (i.e., belonging to the same isomorphism class). 
Conversely, this is not true for the WS and RGG models, which have non-empty neighborhoods independent of the network size. 
Further, for those networks, uniqueness is not strongly influenced by the number of nodes. 
Obviously, if we go to the infinite size, the number of anonymous nodes would increase, but the slope of the uniqueness' boundary is definitely lower than in networks that are locally tree-like. 
Compared to the RGG, the WS model shows a faster transition area from the two extreme values of uniqueness.

The behaviour of the WS changes as $\beta$ changes, (see Fig.~\ref{fig:uniqcurvews_appendix} for uniqueness maps of WS with $\beta$ equal to 0.25 and 0.75). 
If $\beta$ was equal to 1, the figure would have been similar to the one of an ER network, while, if $\beta$ was lower (e.g.~at 0.25, but not close to 0), the uniqueness would be even less dependent on the network size, and the transition between the value of uniqueness from 0 to 1 would have been even faster. 
If $\beta$ was instead 0, then there would not be unique neighborhoods at all, as the WS graph would be in its initial regular ring lattice configuration, where all the nodes have exactly the same neighborhood structure. 

The monotonicity of the uniqueness as a function of network size and average degree, and the two clear distinct areas of zero and full uniqueness that result, imply that we can describe the uniqueness in these models by simple rules. 
However, the two areas are not necessarily comparable in terms of the network anonymization problem, as it is likely that the further we are away from the transition point in the area of full uniqueness, the more changes we would need to make to mitigate an attack's risk using the known anonymization algorithms (see Section \ref{privacy}). 
Too many of these changes make the network very different from the original one, and, consequently, useless for further analysis.

\begin{figure}[h!]
\centering
  \advance\leftskip-0.2cm
  \includegraphics[width=1.0\linewidth]{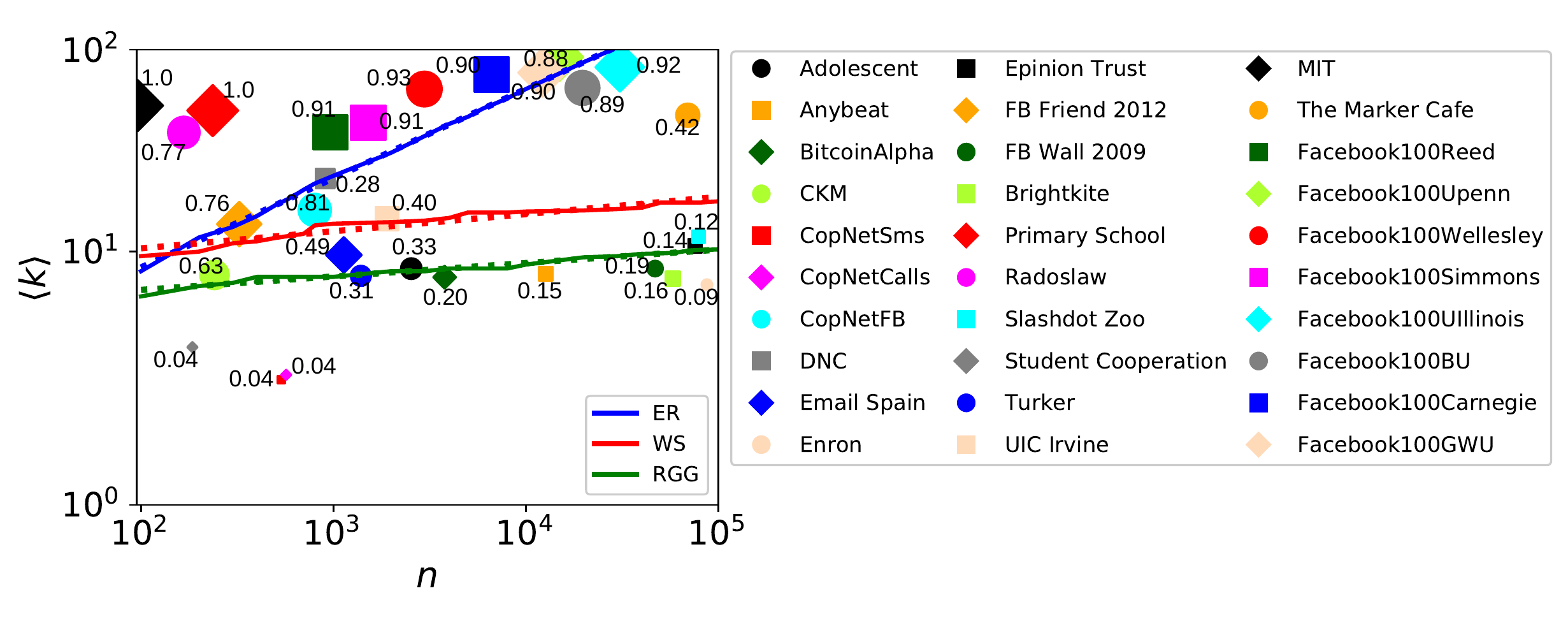}
\caption{Lines representing uniqueness' boundary ($\Uniqneighmath{} = 0.5$) in ER (blue), WS with $\beta = 0.5$(red) and RGG (green) network models, in a log-log scale, and uniqueness of neighborhoods for 30 real-world networks (reported also in Table \ref{tablenets}). The horizontal axis represents the network size $n$, while the vertical axis represents the average degree $\langle k \rangle$. The area below the lines is the one with uniqueness $< 0.5$, while above the lines the uniqueness is $> 0.5$. The continuous lines are the ones computed with the simulations during a binary search process, while the dashed lines are the corresponding linear fits (that have equations $log(y) = m \times log(x) + c$). The dots correspond to the uniqueness of neighborhoods of the real world networks listed on the right, and their size is in according to their value of \Uniqneigh{}. The dots are placed in correspondence to the average degree and network size value of the networks.}
\label{fig:binarysearch}
\end{figure}

The observed monotonicity of the uniqueness allows us to represent the uniqueness of network models using the transition point rather than with the full uniqueness map. 
This also allows us to develop a better computational approach to explore large network sizes. 
To this end we made use of a stochastic version of the binary search algorithm, explained in Appendix \ref{append}, to estimate the uniqueness' boundary, which is the curve corresponding to $\Uniqneighmath = 0.5$, where half of the neighborhoods in the network are uniquely identifiable. 
As Figure \ref{fig:binarysearch} shows, we found that the uniqueness' boundary has a linear trend depending on the network size and average degree. 
This finding implies a simple approximate laws of uniqueness for our networks models, which in turn allows us to predict if networks of given size and density are within the almost fully anonymous or almost fully vulnerable state.
This prediction could serve as a proxy to understand the uniqueness of real-world networks, and to have an idea of how we can modify the network to pass from the unique area to the anonymous one.
In the next section, we discuss how the uniqueness of some real-world networks relates to the ones in the analyzed models, and, based on that, illustrate an alternative strategy to mitigate the re-identification risk from neighborhoods in networks. 

\section{Uniqueness of neighborhoods in empirical networks}\label{uniquenessreal}
In this section, we compare the theoretical results obtained through the analysis of network models, with measurements from 30 empirical social networks (with size varying from 100 to 100000), sampled from the Index of Complex Networks (ICON) \cite{icon}. 
The networks we use in our analysis and the related measures are reported in Table \ref{tablenets}. 
In this table, we also report the expected uniqueness for ER, WSS and RGG models (computed as the mean of 10 realization) with the same size and average degree of each of the 30 empirical networks. 
Fig.~\ref{fig:binarysearch} shows the value of \Uniqneigh{} for those networks, compared to the uniqueness' boundary (\Uniqneigh{} $= 0.5$) of ER, WS and RGG, estimated using binary search. 
Models generally approximate the uniqueness of neighborhoods of empirical networks well, especially the RGG, which is the model with the most pronounced local structure among the ones we analyze. 

The ER model is the worst model in terms of representing the real-world uniqueness' trend. 
Networks that are large and have high average degree, 
such as several of the \textit{Facebook} networks,
have a high uniqueness value, but they are close to the uniqueness boundary point. 
However, the transition between the ``identifiable'' state and the ``anonymous'' one is fast also for ER, thus it can approximately capture high values of uniqueness for smaller \textit{Facebook} networks such as \textit{Penn}, \textit{Carnegie}, \textit{Wellesley} and \textit{Reed}. 
Further, the very small networks \textit{MIT} and \textit{Primary School} have a very high average degree compared to their size, which is also reflected in their relatively high value of uniqueness. 

The WS and RGG models are in most cases better predictors of uniqueness than the ER model. In general, for non-extreme values of uniqueness, the WS model underestimates the uniqueness and RGG overestimates it, so that the truth is somewhere between these two.  
This includes networks like \textit{Epinion Trust}, \textit{Enron}, \textit{Brightkite}, \textit{FB Wall 2009} and \textit{Anybeat}, which are large but
have a low value of \Uniqneigh{}.
They are in the zone of \Uniqneigh{} equal to zero for the ER model and WS is, in some cases, better suited for them. 

There are also some outliers such as \textit{DNC} which is a network where two users are connected if they received the same email, or \textit{The Marker Cafe}, which is an Israeli social network, where one user is connected to another if one is part of the circle of the other. 
Further, in some of the networks, the degree distribution is extremely skewed, some due to sampling where only links incident to certain population are sampled or sampling of one communication channel \cite{torok2016big}. 
This can explain the uniqueness (very low degree nodes are likely to be non-unique). 
The models we analyze here do not take into account skewness of degree distributions.

Obviously, models are not perfect. 
For instance, many of them produce values of uniqueness equal to 1, even when the empirical networks have a slightly lower value. 
The models have indeed a very fast transition between very low value of uniqueness and almost maximum one. 
These extreme values 
may not be present in real-world networks, which have more heterogeneity than our network models. 
In any case, there is an overall pattern that emerges: as predicted by the WS and RGG models, if a network has average degree much higher than 10, there is going to be considerable privacy issues, and if it is much smaller than that then there are almost no privacy issues (at least in terms of uniqueness of neighborhoods) even without an anonymization procedure.

\begin{table}[h!]
\centering
\resizebox{\textwidth}{!}{%
\begin{tabular}{|l|c|c|c|c|c|c|c|c|}
\hline
\multicolumn{1}{|c|}{\textit{\textbf{Net. name}}} & \textit{n} & \textit{m} & $\langle k \rangle$ & \textit{C} & \Uniqneigh & $\langle \Uniqneighmath^{ER} \rangle$ & $\langle \Uniqneighmath^{WS} \rangle$ & $\langle \Uniqneighmath^{RGG} \rangle$ \\ \hline
\textit{Student Cooperation \cite{kumar2016edge}} & 185 & 311 & 3.362 & 0.635 & 0.037 & 0.024 & \textbf{0.035} & 0.11\\ \hline
\textit{CopNetCalls (Copenhagen Network Study - Calls) \cite{sapiezynski2019interaction}} & 568 & 697 & 2.454 & 0.139 & 0.039 & 0.005 & 0.001 & \textbf{0.034} \\ \hline
\textit{CopNetSms (Copenhagen Network Study - Sms) \cite{sapiezynski2019interaction}} & 536 & 621 & 2.317 & 0.155 & 0.044 & 0.006 & 0.002 & \textbf{0.028} \\ \hline
\textit{Enron emails \cite{boldi2011layered}} & 87273 & 299220 & 6.857 & 0.119 & 0.090 & 0.0 & 0.001 & \textbf{0.151} \\ \hline
\textit{Slashdot Zoo \cite{kunegis2009slashdot}} & 79116 & 467731 & 11.823 & 0.058 & 0.119 & 0.0 & \textbf{0.138} & 0.658 \\ \hline
\textit{Epinion Trust \cite{richardson2003trust}} & 75879 & 405740 & 10.694 & 0.137 & 0.143 & 0.0 & \textbf{0.05} & 0.545\\ \hline
\textit{Anybeat \cite{fire2013link}} & 12645 & 49132 & 7.770 & 0.203 & 0.148 & 0.0 & \textbf{0.032} & 0.323\\ \hline
\textit{Brightkite \cite{cho2011friendship}} & 58298 & 214078 & 7.353 & 0.172 & 0.157 & 0.0 & 0.014 & \textbf{0.204} \\ \hline
\textit{FB Wall 2009 (Facebook wall posts) \cite{viswanath2009evolution}} & 46952 & 193494 & 8.242 & 0.107 & 0.193 & 0.0 & 0.015 & \textbf{0.304} \\ \hline
\textit{BitcoinAlpha (trust network) \cite{kumar2016edge}} & 3783 & 14124 & 7.467 & 0.176 & 0.195 & 0.002 & \textbf{0.058} & 0.354\\ \hline
\textit{DNC (Democratic National Committee) emails \cite{kunegis2013konect}} & 906 & 10429 & 23.022 & 0.494 & 0.283 & \textbf{0.544} & 1.0 & 0.98 \\ \hline
\textit{Turker (Amazon Mechanical Turkers) \cite{yin2016communication}} & 1389 & 5267 & 7.585 & 0.291 & 0.306 & 0.01 & 0.107 & \textbf{0.424} \\ \hline
\textit{Adolescent Health survey \cite{moody2001peer}} & 2539 & 10455 & 8.235 & 0.146 & 0.329 & 0.004 & 0.071 & \textbf{0.47} \\ \hline
\textit{UIC Irvine students (Facebook like) \cite{opsahl2009clustering}} & 1899 & 13838 & 14.573 & 0.109 & 0.400 & 0.025 & \textbf{0.654} & 0.919 \\ \hline
\textit{The Marker Cafe \cite{fire2014computationally}} & 69413 & 1644849 & 47.393 & 0.186 & 0.424 & \textbf{0.002} & 1.0 & 0.988 \\ \hline
\textit{Email Spain (Universitat Rovira i Virgili) \cite{guimera2003self}} & 1133 & 5451 & 9.622 & 0.220 & 0.492 & 0.02 & 0.292 & \textbf{0.659} \\ \hline
\textit{CKM physician social network \cite{burt1987social}} & 241 & 924 & 7.668 & 0.311 & 0.634 & 0.11 & 0.295 & \textbf{0.528}\\ \hline
\textit{FB friends 2012 (Facebook friendships) \cite{blagus2012self}} & 324 & 2218 & 13.691 & 0.465 & 0.762 & \textbf{0.422} & 0.952 & 0.905 \\ \hline
\textit{Radoslaw (Manufacturing company emails) \cite{kumar2016edge}} & 167 & 3251 & 38.934 & 0.591 & 0.766 & \textbf{1.0} & \textbf{1.0} & \textbf{1.0} \\ \hline
\textit{CopNetFB (Copenhagen Network Study - Facebook) \cite{sapiezynski2019interaction}} & 800 & 6429 & 16.072 & 0.315 & 0.81 & 0.175 & \textbf{0.945} & 0.948 \\ \hline
\textit{Facebook100UPenn \cite{traud2012social}} & 14916 & 686501 & 92.048 & 0.213 & 0.882 & \textbf{0.926} & 1.0 & 1.0 \\ \hline
\textit{Facebook100BU \cite{traud2012social}} & 19700 & 637528 & 64.723 & 0.190 & 0.886 & 0.107 & 1.0 & \textbf{0.998} \\ \hline
\textit{Facebook100Carnegie \cite{traud2012social}} & 6637 & 249967 & 75.325 & 0.278 & 0.896 & \textbf{0.997} & 1.0 & 1.0 \\ \hline
\textit{Facebook100GWU \cite{traud2012social}} & 12193 & 469528 & 77.015 & 0.217 & 0.903 & 0.747 & 1.0 & \textbf{0.999}\\ \hline
\textit{Facebook100Reed \cite{traud2012social}} & 963 & 18812 & 39.069 & 0.318 & 0.906 & 1.0 & 1.0 & \textbf{0.997} \\ \hline
\textit{Facebook100Simmons \cite{traud2012social}} & 1518 & 32988 & 43.462 & 0.314 & 0.907 & 0.999 & 1.0 & \textbf{0.998} \\ \hline
\textit{Facebook100UIllinois \cite{traud2012social}} & 30809 & 1264828 & 82.081 & 0.214 & 0.920 & 0.127 & 1.0 & \textbf{0.999} \\ \hline
\textit{Facebook100Wellesley \cite{traud2012social}} & 2970 & 94899 & 63.905 & 0.264 & 0.932 & 1.0 & 1.0 & \textbf{0.999} \\ \hline
\textit{MIT \cite{konect:2017:mit}} & 96 & 2539 & 52.895 & 0.751 & 1.0 & \textbf{1.0} & \textbf{1.0} & \textbf{1.0} \\ \hline
\textit{Primary School dynamic contacts \cite{stehle2011high}} & 236 & 5899 & 49.991 & 0.501 & 1.0 & \textbf{1.0} & \textbf{1.0} & 0.999 \\ \hline
\end{tabular}
}

\caption{Datasets' basic measures (number of nodes $n$, number of edges $m$, average degree $\langle k \rangle$, average local clustering coefficient $C$), uniqueness of neighborhoods \Uniqneigh{}, and expected \Uniqneigh{} values for the corresponding ER, WS and RGG models (with the same number of nodes and average degree), computed as a mean of 10 realizations. The table is sorted by increasing value of \Uniqneigh{}. The models' expected \Uniqneigh{} values that are closer to each real network are highlighted in bold (if two models produced the same value, they are both highlighted).}
\label{tablenets}
\end{table}

\section{Mitigating neighborhood attacks}\label{anonymizingnet}
In the previous section, we have seen that, in terms of understanding privacy, models can be a good approximation for real-world social network data, and that, typically, social networks are better approximated by models with a clustered local structure, such as the RGG. 
Ultimately, we are interested in sharing network data while protecting user's privacy or, at least, mitigating the risk of identity disclosure, while keeping the data useful for exploration and analysis. 
Due to the presence of two distinct areas with the almost maximum and minimum values of \Uniqneigh{}, we can first, due to the models, form an idea of ``how far" a network is from the area where all the nodes are anonymous ($\Uniqneighmath \approx 0$). 
This knowledge can, depending on the anonymization strategy, also help in understanding how many modifications we need to make the network anonymous. 
We can perform these estimations even before we have the full data available just by having an estimate on the network size and density.

Inspired by the strong relationship between average degree and uniqueness, we now investigate a simple strategy to mitigate neighborhood attack based on lowering the average degree by random edge sampling. 
At first this trivial method might seem inferior to the methods that try to minimize the number of changes to the network. 
However, there are two main advantages of sampling: 
first, sampling is extremely simple compared to the optimisation heuristics, making it practical and even possible to use already at the data collection phase, even before collecting the whole network (which would be impossible with the current optimisation methods); 
second, random sampling produces statistically tractable changes to the network -- as opposed to optimisation algorithms for which one in practise needs to assume that the changes are in the worst possible places.

As we discuss below, the application of this method need to be coupled with analysis of also other risk factors prior to data sharing (e.g. the presence of additional indicators or other public datasets which could lead the re-identification of a node by other means), but the method is simple, intuitive and practical, as it could potentially be applied already in the data collection phase. 

\subsection{Random sampling of edges}\label{randomsampling}
By exploiting the fact that the uniqueness in real-world social networks is heavily influenced by the average degree, we propose a method to mitigate neighborhood attack in social networks with respect to uniqueness of neighborhoods by randomly sampling edges. 
This methods leaves the number of nodes unchanged, while lowering the average degree and, consequently, lowering the uniqueness of neighborhoods. 
Random edge sampling makes 
more statistical guarantees on the final analysis, allowing the estimation of original measures through statistical methods \cite{Kolaczyk2009Statistical}. 
Conversely, with existing methods, reversing the effect of the anonymization algorithm could be difficult.  

The disadvantages of random edge sampling is that each piece of information that is shared is actually true. 
That means, each published edge actually exists, revealing real relationships between the entities represented. 
In addition, 
a careful analysis of risk deriving from the presence of possible nodes' attributes and node identities would need to be carried out, in order to decide the best anonymization or pseudo-anonymization method. 
However, if the nodes have no attributes or no unique-identifier attributes, the proposed method might suit the case. 
In the presence of attributes, in any case, the computation of uniqueness should be carried out taking into account those as well. 
In this paper, we focus only on structural uniqueness, without considering any nodes' attribute.

\begin{figure}[h!]
\centering
  \advance\leftskip-0.2cm
  \includegraphics[]{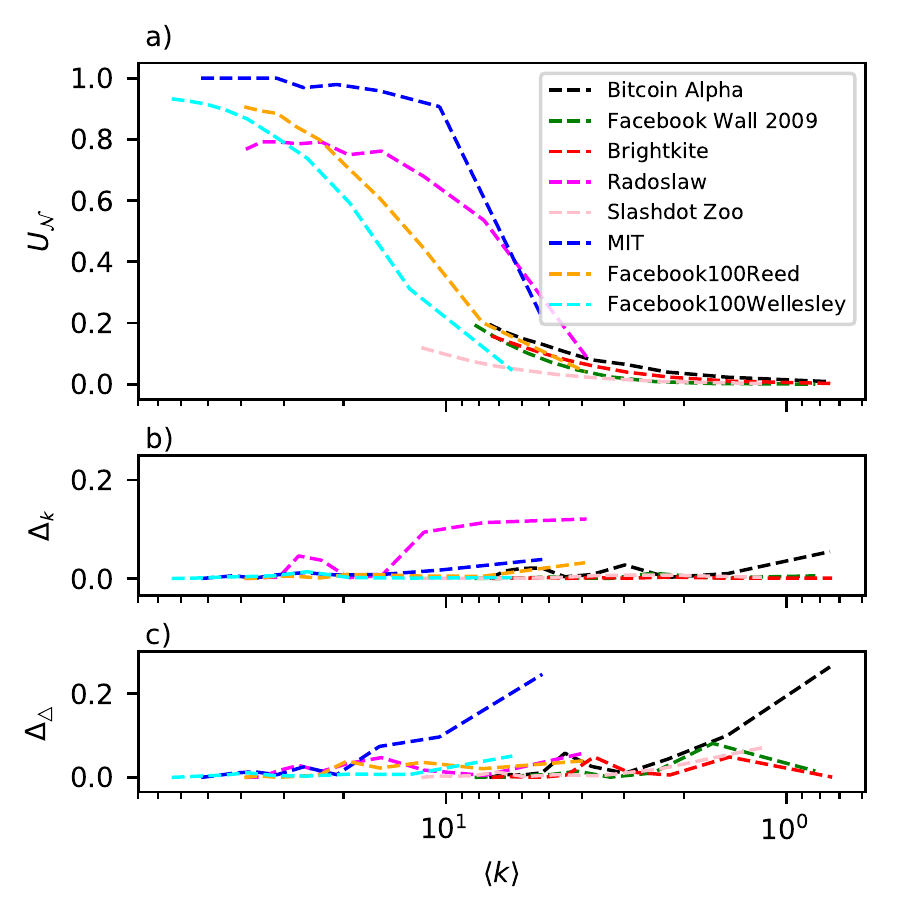}
\caption{Results of uniform random edge-sampling (with decreasing sampling ratio from 1.0 to 0.1) on the uniqueness of neighborhoods of 8 real networks (indicated in the legend at the top-right), and error of the estimation of network measures after sampling. The features of the considered 8 networks can be read in Table \ref{tablenets}. The horizontal axis represents the average degree of the networks after sampling. The curves start from the average degree value of the original network. (\textit{a}) Effect of uniform edge sampling on the value of \Uniqneigh (on the vertical axis): as we sample edges, the values of the average degree $\langle k \rangle$ gets lower as well; (\textit{b}) Error of the estimation of the degree of the nodes (average over all the nodes in the network) in the networks with Eq.~\ref{eq:degestimation}; (\textit{c}) Error of the estimation of the amount of triangles in the networks with Eq.~\ref{eq:trestimation}. The error in these estimates is the difference, in absolute value, between the original value and the estimated one.}
\label{fig:anonymiz_net}
\end{figure}

Fig.~\ref{fig:anonymiz_net} shows the effect of the random sampling of edges on the average degree and uniqueness of neighborhoods. In general, the higher the amount of sampling we perform, the lower the numerical value of those two measures would be. We report results for 8 of the networks we listed in Table \ref{tablenets}. We chose those networks to have different size, average degree, and value of uniqueness.

We can see in the figure that the real-world networks follow the same pattern we observed for our model networks in Section \ref{uniquenessmaps}: the uniqueness monotonically decreases as the average degree decreases. 
Further, there is a transition-like behavior close to the uniqueness boundary, such that networks that have a high original value of uniqueness ($\approx 1.0$) first experience a slow decrease after the initial samplings and, then, a very rapid fall in the uniqueness values, until they get more stable again towards the lowest values of uniqueness (i.e., with the highest sampling rates). 
In fact, it seems that, when sampled, networks display a rapid transition (represented by the rapid decreasing of the uniqueness' value) between a state where nodes are almost all unique and a state where there is no unique node. 
In models, the area in between the two extreme zones was also relatively small, meaning that the range of average degree needed to pass from the ``identifiable'' state to the ``anonymous" state is relatively narrow.

As we sample edges, we are modifying the neighborhoods at random, introducing uncertainty. 
For this reason, in real-world applications, we would not necessarily need to sample until we reach very low values of uniqueness, as this could significantly modify the network. 
Algorithms in the literature aim to reach uniqueness values of zero, but, even with their methods, once a neighborhood is modified by, for instance, adding one edge, it is already different from the original, lowering the attacker's success probability to identify it. 
In our case, we remove edges from the neighborhood and, after some samples, we could stop, in order to preserve more edges and to avoid lowering the utility too excessively. 
To be more precise on this point, we would need to perform a further study on ``how far" the neighborhoods in the network are from the original one and from other neighborhoods. 
This analysis would require measuring their edit-distance and the actual risk of a neighborhood to be re-identified. 
Such an analysis is, however, beyond the scope of our study. 

\subsection{Error evaluation and correction}\label{erreval}
As we are sampling edges uniformly, we can reconstruct measures of the original network, given the sampling rate. 
In this section, we show that data anonymized via our method allows the end user to make unbiased estimates and related error estimation of network statistics by employing network sampling theory \cite{Kolaczyk2009Statistical}. 
In Fig.~\ref{fig:anonymiz_net}b-c we show the error estimation of two basic network measures, degree of the nodes, and number of triangles, derived from the sampling of edges in the 8 networks reported in Fig.~\ref{fig:anonymiz_net}a.

Each node $i$'s degree $k_{i}$, after sampling, can be estimated with the following formula:
\begin{equation}\label{eq:degestimation}
    \widehat{k}_{i} = \frac{k_{i}^{s}}{s}  \:;
\end{equation}
where $s$ is the sampling rate and $k_{i}^{s}$ is the degree of node $i$ observed in the network after sampling. 
The amount of triangles $\bigtriangleup$ in the network can be estimated with: 
\begin{equation}\label{eq:trestimation}
    \widehat{\bigtriangleup} = \frac{\bigtriangleup^{s}}{s^{3}}\,,
\end{equation}
where $\bigtriangleup^{s}$ is the amount of triangles observed in the network after sampling. 
The error in these estimates is the difference, in absolute value, between the original value and the estimated value. 
For each sampling rate and each network considered, we show the error of the estimation in Fig.~\ref{fig:anonymiz_net}b-c. 
In terms of degree error, we show the average error over all nodes. 
We see that the error is relatively small for even relatively high values of sampling (and, consequently, low values of average degree), especially for the degree. 
The error is a bit higher (slightly above 0.2), for the triangle-count. 
The latter is a slightly more complicated measure than the degree, so it is expected that, with higher sampling rate, we obtain an higher estimation error. 
However, as we mentioned before in subsection \ref{randomsampling}, we could not necessarily need to sample until we reach the minimum uniqueness. 
In any case, the average errors on these networks are still quite low, confirming that our method can give some statistical guarantees on any measure, allowing at least a certain confidence in estimating statistics from the original network.

\section{Conclusion and discussion}\label{conclusion}
In this paper we studied the uniqueness of neighborhoods in networks, discovering that there are regularities in the uniqueness behaviour of network models, by investigating its dependencies on average degree and network size. 
We have seen that, in models, there is a linear uniqueness' boundary which separates two zones where nodes are either almost fully vulnerable to identification ($\Uniqneighmath \approx 1.0$) or almost fully anonymous ($\Uniqneighmath \approx 0$).
The more the network is locally dense, the less the uniqueness depends on the network size, but is mostly affected by the average degree. 
We have seen that these findings hold for many real-world social networks, which are well approximated by models such as the RGG. 
Inspired by this finding, we have shown that randomly sampling edges can be an effective method to lower the identity disclosure risk. 
This method provides statistical guarantees when analyzing the shared data, an advantage compared to the existing algorithms, which are optimized for specific purposes and where the effect of the anonymization on the network statistics could be hard to understand and reverse. 

A number of questions still remains open. 
For example, we could also study what happens in the ``unique" zone of a network, if the average degree keeps growing. 
As the networks move further away from the uniqueness' boundary, and the uniqueness reaches its maximum value, we do not know whether the edit distance between the neighborhoods increases or remains stable. 
Even though we would expect it to increase, a further study is needed to confirm.  
This understanding would be useful for more accurately estimating if a network is ``too far'' from the anonymous area, meaning that lowering the uniqueness of neighborhoods via any anonymization process would lower the utility in a radical way, making the final analysis much less useful. 

Our edge-sampling method introduces uncertainty even before the resulting uniqueness of the full network is zero, as it modifies the neighborhoods anyway. 
It could therefore be possible to sample less edges. 
The current algorithms run until they reach uniqueness equal to zero, introducing probably more modifications than needed. 
To better understand when to stop introducing changes, we could study the best guess of an attacker (for instance by computing the edit distance between the target neighborhood and the existing ones) in re-identifying a certain target node after the network has been anonymized with existing methods or sampled with uniform edge-sampling.

Random edge sampling would work as it is a simple method to implement already in advance, for instance during the data collection phase (e.g. removing links before they get stored at all or using the uniqueness maps to have an estimation of the dimension of the study before conducting it), but it is worth mentioning that this method also comes with obvious limitations. 
For instance, 
links are still fully preserved (revealing actual relationships) and only true information from the original network is shown. 
Moreover, our uniqueness' study is carried out on networks without any node or edge attributes. 
This types of networks are of interests when studying social networks from a structural point of view, where a nodes' neighborhood could be sensitive from a privacy point of view for revealing, for instance, the presence of additional relationships or nodes' attribute which should be kept private. 
However, if nodal feature are present, those may also be taken into account in the computation of uniqueness, as they could act as identifiers, and could potentially help the attacker to identify targets. 

The current trend of research for privacy-preserving data sharing is moving towards synthetic data generation, for example applying generative models with differential privacy \cite{sala2011sharing, xiao2014differentially}. 
This line of methods have some clear advantages over sharing modified networks. 
Differential privacy can guarantee certain amount of privacy against any kind of attack scenario, whereas we only studied neighborhood attack. 
One could also have studied node automorphisms or other stricter notions of privacy, but meeting these guarantees for large networks can be practically impossible. 
On the other hand, synthetic data generation is only as useful as the model behind it for a specific research question. 
For example, stochastic block models are unable to capture many important aspects of local structure and $dK$-series do not adequately model information on the mesoscopic structures. In general, using model generated data limits its usability to questions that can be answered by analysing the model.
Data anonymization is a compromise between utility and anonymity \cite{horawalavithana2019privacy, shokri2015privacy}. 

As we have seen that the uniqueness of neighborhoods is affected by the network configuration, further studies can be conducted on different types of networks, such as multiplex networks, to understand how the uniqueness is affected when the attacker has multiple data sources at its disposal.  
One can further analyse additional and more complicated models, that take into consideration specific nature of the network data and processes generating it. 

In general, networks are difficult to anonymize and neighborhoods are certainly not the only aspect to take into account when considering the privacy risk in data sharing. 
However, structural privacy is an additional point that should be certainly considered, and neighborhoods may be a crucial threat for identifying a certain entity, especially if the attacker has domain-knowledge and sensitive attributes are shared with the nodes (even if nodes' identity are dropped). 
Our study suggests a method to lower this risk, while keeping data useful for general statistical analysis, and revealing new regularities for network models, as well as possibly suggesting a methodology to further investigate this and other structural properties of networks and their link with privacy and, especially, identity disclosure risk.

Random network models are in general useful for two purposes: for theoretical understanding of the underlying phenomena and as proxies for real networks. 
For example, spreading processes on networks are routinely and successfully analysed using random network models.
Here we have explored these two purposes in the context of network anonymization. 
Our findings bring theoretical understanding to the factors making some networks vulnerable to neighborhood attack and others not. 
Further, they allow one to estimate the risk even without seeing the network data. 
This aspect can be useful for example before data collection, to justify why a network dataset is sensitive without sharing it, or when deciding if it is even worth to apply network anonymization to it.

\appendix
\section{Appendix: Additional figures}\label{appendplots}

In this appendix, we show additional plots regarding the uniqueness of neighborhoods and degree uniqueness in network models, in order to let the reader have a more complete picture of the behaviour of such models. 

\begin{figure}[h!]
\centering
  \advance\leftskip-0.2cm
  \includegraphics[]{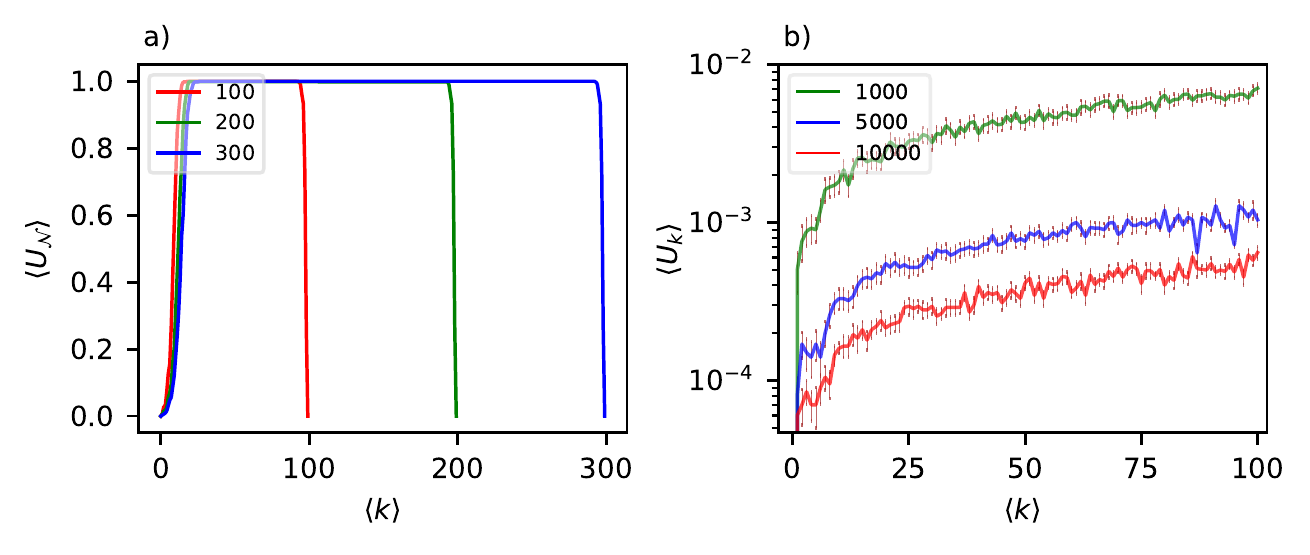}
\caption{Neighborhood uniqueness and degree uniqueness in ER networks.
    (\textit{a}) The expected uniqueness of neighborhoods $\langle U_{\mathcal{N}} \rangle$ (in a linear scale), in ER networks of size 100, 200, 300 (computed as the mean of 10 independent network realizations), for for the complete range of average degree (from 0 to $n-1$). (\textit{b}) Expected degree uniqueness $\langle U_{k} \rangle$  (in a log scale) for networks of size 1000, 5000 and 10000, for a range of average degree $\langle k \rangle$ from 0 to 100. The error bar represents the standard error of the mean. The means and errors are computed over 20 independently simulated networks.}\label{fig:fulluniqcurve}
\end{figure}

Fig.~\ref{fig:fulluniqcurve}a shows the uniqueness curve for the complete range of average degree (from 0 to $n-1$) for ER networks of size 100, 200, 300. The values of uniqueness are computed as a mean over 10 realization of each ER configuration. We can see that the value of uniqueness goes to one (maximum value) almost immediately for all the three networks, and it experiences a dramatic fall only towards the real end, when the network is almost complete (that is, every node is connected to any other). 

Fig.~\ref{fig:fulluniqcurve}b shows instead the expected degree uniqueness for ER networks of 1000, 5000 and 10000, for a range of average degree from 0 to 100. Those curves are also shown in Fig.~\ref{fig:ER_uniqueness}a on a linear scale, where they were covered by each other because of the very low values. In Fig.~\ref{fig:fulluniqcurve}b we use a log scale instead, in order to make those curves more visible and have an idea on how bigger networks such as the ones of 5000 and 10000 nodes compare to the smaller one with 1000, already shown also in Fig.~\ref{fig:ER_uniqueness}b along with the degree uniqueness curve for ER networks with 100 nodes. In this case, the values of the degree uniqueness and error bar (which represents the standard error of the mean), are computed as a mean of 20 realizations.

\begin{figure}[h!]
\centering
  \advance\leftskip-0.2cm
  \includegraphics[]{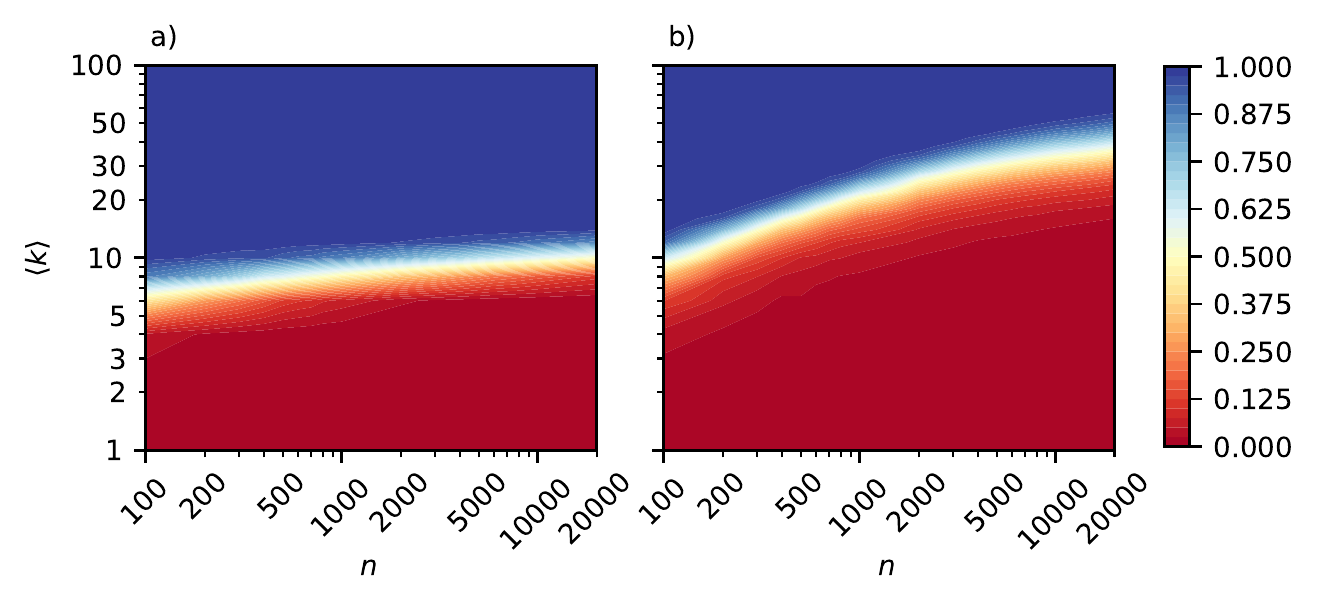}
\caption{Uniqueness maps of WS networks with probability of rewiring $\beta = 0.25$ (\textit{a}) and $\beta = 0.75$ (\textit{b}). The heatmaps represent the variation of the uniqueness of neighborhoods value (in color: the blue area corresponds to a uniqueness value equal to 1, while the red area corresponds to a uniqueness value equal to 0), in function of the average degree $\langle k \rangle$ (on the vertical axis) and network size $n$ (on the horizontal axis). The axis are in logarithmic scale. The uniqueness values are computed as the mean of 10 independently simulated networks for each average degree and network size value. Note that, even if not clearly shown in this figure, in networks with 100 nodes, the maximum possible value of average degree is 99, where $\Uniqneighmath = 0$ (see Fig.~\ref{fig:ER_uniqueness} or Fig.~\ref{fig:fulluniqcurve} for an example of the uniqueness behaviour in ER networks with 100 nodes).}\label{fig:uniqcurvews_appendix}
\end{figure}

Fig.~\ref{fig:uniqcurvews_appendix} shows the uniqueness maps for WS networks with probability of rewiring $\beta$ of 0.25 ((a) and 0.75 (b). The higher the probability of rewiring, the closer the WS network would be to a random network, like the ER. In fact, we can see that the map of WS with $\beta = 0.75$ is close to the map of the ER network in Fig.~\ref{fig:uniquenessmaps}a. It would be almost the same at the extreme, with $\beta = 1$ (i.e. all the edges are rewired, realizing, in practice, random network). As we observed that the uniqueness' boundary has a linear trend, we can see that this similarity is reflected by the slope of the $\langle \Uniqneighmath \rangle = 0.5$ curve. On the other hand, when $\beta = 0.25$, the network is much more dense, consequently the uniqueness' behaviour is more similar to the RGG model, shown in Fig.~\ref{fig:uniquenessmaps}c. However, if $\beta = 0$, the network would be a regular lattice (as per the initial configuration of the WS model), thus the neighborhoods of all nodes would be isomorphic to each other, and the network would have a uniqueness of neighborhoods' value equal to zero. The slope of the WS network with $\beta = 0.5$ (Fig~\ref{fig:uniquenessmaps}b) is, as expected, in between the ones with $\beta$ to 0.25 and 0.75.

\section{Appendix: Stochastic binary search algorithm}\label{append}
In this appendix, we explain the modification to the binary search algorithm we use in subsection \ref{uniquenessmaps}. We run the binary search to estimate the uniqueness boundary ($\Uniqneighmath=0.5$) in network models and better compare \Uniqneigh in different models. The algorithm looks for a certain value of uniqueness in a network generated according to a model with a given number of nodes, in a range of average degree values delimited by two extremes. 

Our algorithm is a stochastic and continuous version of the classic binary search \cite{weissteinbinary}. The algorithm searches for a target value (in our case, $0.5$) in a certain range, evaluating first the extreme values of the interval and, if those are not the ones we are looking for, evaluating the middle value. If the middle value corresponds to the target value, the algorithm stops, otherwise it continues the search process recursively, until it finds the target value. The interval used in the new evaluation corresponds to either the lower part of the interval (delimited by the original lower extreme and the middle value), or the upper part (delimited by the middle value and the original upper extreme). The algorithm stops also if the extremes of the interval we are evaluating are too close to each other.

In our case, to decide on which side of the interval to move, we exploit the fact that, with a fixed network size, we know from the simulations that the uniqueness grows with the average degree (at least in the sparse region). Consequently, if a certain average degree value we are evaluating gives a uniqueness value higher than the one we want, we move to the left (or lower) part of the interval, which is the one containing lower values; otherwise we move to the right (or higher) part, by always computing the middle value of the new interval. 

To compute the uniqueness value corresponding to each average degree, we generate five networks with that average degree and the given network size, and we take the mean of the corresponding uniqueness values. Since we want certain guarantees every time we decide which new interval to evaluate (and also when to stop), we compute a confidence interval (at $99\%$ confidence level) of the mean of the uniqueness value of the networks generated with certain parameters. We then check whether the target uniqueness value is contained in that interval: if it is not, we move either to the right or the left side; if it is, we run new simulations to have a better estimation of the real mean. If the target uniqueness value is still in the interval after a maximum number of simulation (we chose 30), then we can be confident that the evaluated average degree is the one we are looking for. Otherwise, we continue with the search. We also set a 0.02 tolerance level to the target uniqueness value. That means, if we find an average degree corresponding to either $0.52$ or $0.48$, the search process is considered successfully ended.

\bibliographystyle{unsrt}  
\bibliography{references}  


\end{document}